\documentclass[12pt]{article}
\usepackage{epsfig}
\usepackage{amsmath}
\usepackage{hhline}
\usepackage{amssymb}
\usepackage{times}
\usepackage{cite}

\newlength{\dinwidth}
\newlength{\dinmargin}
\begin{document}  
\newcommand{\pom}{{I\!\!P}}
\newcommand{\reg}{{I\!\!R}}
\newcommand{\slowpi}{\pi_{\mathit{slow}}}
\newcommand{\fiidiii}{F_2^{D(3)}}
\newcommand{\fiidiiiarg}{\fiidiii\,(\beta,\,Q^2,\,x)}
\newcommand{\n}{1.19\pm 0.06 (stat.) \pm0.07 (syst.)}
\newcommand{\nz}{1.30\pm 0.08 (stat.)^{+0.08}_{-0.14} (syst.)}
\newcommand{\fiidiiiful}{F_2^{D(4)}\,(\beta,\,Q^2,\,x,\,t)}
\newcommand{\fiipom}{\tilde F_2^D}
\newcommand{\ALPHA}{1.10\pm0.03 (stat.) \pm0.04 (syst.)}
\newcommand{\ALPHAZ}{1.15\pm0.04 (stat.)^{+0.04}_{-0.07} (syst.)}
\newcommand{\fiipomarg}{\fiipom\,(\beta,\,Q^2)}
\newcommand{\pomflux}{f_{\pom / p}}
\newcommand{\nxpom}{1.19\pm 0.06 (stat.) \pm0.07 (syst.)}
\newcommand {\gapprox}
   {\raisebox{-0.7ex}{$\stackrel {\textstyle>}{\sim}$}}
\newcommand {\lapprox}
   {\raisebox{-0.7ex}{$\stackrel {\textstyle<}{\sim}$}}
\def\gsim{\,\lower.25ex\hbox{$\scriptstyle\sim$}\kern-1.30ex%
\raise 0.55ex\hbox{$\scriptstyle >$}\,}
\def\lsim{\,\lower.25ex\hbox{$\scriptstyle\sim$}\kern-1.30ex%
\raise 0.55ex\hbox{$\scriptstyle <$}\,}
\newcommand{\pomfluxarg}{f_{\pom / p}\,(x_\pom)}
\newcommand{\dsf}{\mbox{$F_2^{D(3)}$}}
\newcommand{\dsfva}{\mbox{$F_2^{D(3)}(\beta,Q^2,x_{I\!\!P})$}}
\newcommand{\dsfvb}{\mbox{$F_2^{D(3)}(\beta,Q^2,x)$}}
\newcommand{\dsfpom}{$F_2^{I\!\!P}$}
\newcommand{\gap}{\stackrel{>}{\sim}}
\newcommand{\lap}{\stackrel{<}{\sim}}
\newcommand{\fem}{$F_2^{em}$}
\newcommand{\tsnmp}{$\tilde{\sigma}_{NC}(e^{\mp})$}
\newcommand{\tsnm}{$\tilde{\sigma}_{NC}(e^-)$}
\newcommand{\tsnp}{$\tilde{\sigma}_{NC}(e^+)$}
\newcommand{\st}{$\star$}
\newcommand{\sst}{$\star \star$}
\newcommand{\ssst}{$\star \star \star$}
\newcommand{\sssst}{$\star \star \star \star$}
\newcommand{\tw}{\theta_W}
\newcommand{\sw}{\sin{\theta_W}}
\newcommand{\cw}{\cos{\theta_W}}
\newcommand{\sww}{\sin^2{\theta_W}}
\newcommand{\cww}{\cos^2{\theta_W}}
\newcommand{\trm}{m_{\perp}}
\newcommand{\trp}{p_{\perp}}
\newcommand{\trmm}{m_{\perp}^2}
\newcommand{\trpp}{p_{\perp}^2}
\newcommand{\alp}{\alpha_s}

\newcommand{\alps}{\alpha_s}
\newcommand{\sqrts}{$\sqrt{s}$}
\newcommand{\LO}{$O(\alpha_s^0)$}
\newcommand{\Oa}{$O(\alpha_s)$}
\newcommand{\Oaa}{$O(\alpha_s^2)$}
\newcommand{\PT}{p_{\perp}}
\newcommand{\JPSI}{J/\psi}
\newcommand{\sh}{\hat{s}}
\newcommand{\uh}{\hat{u}}
\newcommand{\MP}{m_{J/\psi}}
\newcommand{\PO}{I\!\!P}
\newcommand{\xbj}{x}
\newcommand{\xpom}{x_{\PO}}
\newcommand{\ttbs}{\char'134}
\newcommand{\xpomlo}{3\times10^{-4}}  
\newcommand{\xpomup}{0.05}  
\newcommand{\dgr}{^\circ}
\newcommand{\pbarnt}{\,\mbox{{\rm pb$^{-1}$}}}
\newcommand{\gev}{\,\mbox{GeV}}
\newcommand{\WBoson}{\mbox{$W$}}
\newcommand{\fbarn}{\,\mbox{{\rm fb}}}
\newcommand{\fbarnt}{\,\mbox{{\rm fb$^{-1}$}}}
\newcommand{\dsdx}[1]{$d\sigma\!/\!d #1\,$}
\newcommand{\eV}{\mbox{e\hspace{-0.08em}V}}
%
%
\newcommand{\qsq}{\ensuremath{Q^2} }
\newcommand{\gevsq}{\ensuremath{\mathrm{GeV}^2} }
\newcommand{\et}{\ensuremath{E_t^*} }
\newcommand{\rap}{\ensuremath{\eta^*} }
\newcommand{\gp}{\ensuremath{\gamma^*}p }
\newcommand{\dsiget}{\ensuremath{{\rm d}\sigma_{ep}/{\rm d}E_t^*} }
\newcommand{\dsigrap}{\ensuremath{{\rm d}\sigma_{ep}/{\rm d}\eta^*} }

\newcommand{\dstar}{\ensuremath{D^*}}
\newcommand{\dstarp}{\ensuremath{D^{*+}}}
\newcommand{\dstarm}{\ensuremath{D^{*-}}}
\newcommand{\dstarpm}{\ensuremath{D^{*\pm}}}
\newcommand{\zDs}{\ensuremath{z(\dstar )}}
\newcommand{\Wgp}{\ensuremath{W_{\gamma p}}}
\newcommand{\ptds}{\ensuremath{p_t(\dstar )}}
\newcommand{\etads}{\ensuremath{\eta(\dstar )}}
\newcommand{\ptj}{\ensuremath{p_t(\mbox{jet})}}
\newcommand{\ptjn}[1]{\ensuremath{p_t(\mbox{jet$_{#1}$})}}
\newcommand{\etaj}{\ensuremath{\eta(\mbox{jet})}}
\newcommand{\detadsj}{\ensuremath{\eta(\dstar )\, \mbox{-}\, \etaj}}

\def\Journal#1#2#3#4{{#1} {\bf #2} (#3) #4}
\def\NCA{\em Nuovo Cimento}
\def\NIM{\em Nucl. Instrum. Methods}
\def\NIMA{{\em Nucl. Instrum. Methods} {\bf A}}
\def\NPB{{\em Nucl. Phys.}   {\bf B}}
\def\PLB{{\em Phys. Lett.}   {\bf B}}
\def\PRL{\em Phys. Rev. Lett.}
\def\PRD{{\em Phys. Rev.}    {\bf D}}
\def\ZPC{{\em Z. Phys.}      {\bf C}}
\def\EJC{{\em Eur. Phys. J.} {\bf C}}
\def\CPC{\em Comp. Phys. Commun.}

\newcommand{\be}{\begin{equation}} 
\newcommand{\ee}{\end{equation}} 
\newcommand{\ba}{\begin{eqnarray}} 
\newcommand{\ea}{\end{eqnarray}}

\begin{titlepage}

\noindent
\begin{flushleft}
{\tt DESY 08-065    \hfill    ISSN 0418-9833} \\
{\tt June 2008}                  \\
\end{flushleft}

\vspace{2cm}
\begin{center}
\begin{Large}

{\bf Multi--Lepton Production at High Transverse Momenta in \begin{boldmath}$ep$\end{boldmath} Collisions at HERA \\}

\vspace{2cm}

H1 Collaboration

\end{Large}
\end{center}

\vspace{2cm}

\begin{abstract}

Processes leading to a final state with at least two high transverse momentum leptons (electrons or muons) are studied using the full $e^\pm p$ data sample collected by the H1 experiment at HERA.
The data correspond to an integrated luminosity of $463$~pb$^{-1}$.
Di-lepton and tri-lepton event classes are investigated.
Cross sections of the production of $e^+e^-$ and $\mu^+\mu^-$ pairs are derived in a restricted phase space dominated by photon-photon collisions.
In general, good agreement is found with Standard Model predictions.
Events are observed with a total scalar sum of lepton transverse momenta above $100$~GeV where the Standard Model expectation is low.
In this region, combining di-lepton and tri-lepton classes, five events are observed in $e^+p$ collisions, compared to a Standard Model expectation of \mbox{$0.96 \pm 0.12$}, while no such event is observed in $e^-p$ data for $0.64 \pm 0.09$ expected.

\end{abstract}

\vspace{1cm}

\begin{center}
Dedicated to the memory of our dear friend and colleague, Beate Naroska.
\end{center}

\vspace{0.5cm}

\begin{center}
Submitted to \PLB
\end{center}

\end{titlepage}

%
%
%
\begin{flushleft}

F.D.~Aaron$^{5,49}$,           
C.~Alexa$^{5}$,                
V.~Andreev$^{25}$,             
B.~Antunovic$^{11}$,           
S.~Aplin$^{11}$,               
A.~Asmone$^{33}$,              
A.~Astvatsatourov$^{4}$,       
A.~Bacchetta$^{11}$,           
S.~Backovic$^{30}$,            
A.~Baghdasaryan$^{38}$
% ...

\bigskip{\it
 $ ^{1}$ I. Physikalisches Institut der RWTH, Aachen, Germany$^{ a}$ \\
% ...
$ ^{4}$ Inter-University Institute for High Energies ULB-VUB, Brussels;
          Universiteit Antwerpen, Antwerpen; Belgium$^{ c}$ \\
 $ ^{5}$ National Institute for Physics and Nuclear Engineering (NIPNE) ,
          Bucharest, Romania \\
% ...
$ ^{11}$ DESY, Hamburg, Germany \\
 $ ^{12}$ Institut f\"ur Experimentalphysik, Universit\"at Hamburg,
          Hamburg, Germany$^{ a}$ \\
% ...
$ ^{25}$ Lebedev Physical Institute, Moscow, Russia$^{ e}$ \\
% ...
$ ^{30}$ Faculty of Science, University of Montenegro,
          Podgorica, Montenegro$^{ e}$ \\
% ...
$ ^{33}$ Dipartimento di Fisica Universit\`a di Roma Tre
          and INFN Roma~3, Roma, Italy \\
% ...
$ ^{38}$ Yerevan Physics Institute, Yerevan, Armenia \\
% ...
$ ^{49}$ Also at Faculty of Physics, University of Bucharest,
          Bucharest, Romania \\
% ...
$ ^a$ Supported by the Bundesministerium f\"ur Bildung und Forschung, FRG,
      under contract numbers 05 H1 1GUA /1, 05 H1 1PAA /1, 05 H1 1PAB /9,
      05 H1 1PEA /6, 05 H1 1VHA /7 and 05 H1 1VHB /5 \\
% ...
$ ^c$ Supported by FNRS-FWO-Vlaanderen, IISN-IIKW and IWT
      and  by Interuniversity
Attraction Poles Programme,
      Belgian Science Policy \\
% ...
$ ^e$ Supported by the Deutsche Forschungsgemeinschaft \\
% ...
}\end{flushleft}
%

\newpage

\section{Introduction}

Within the Standard Model (SM) the production of multi-lepton events
in electron\footnote{In this paper the term ``electron'' is used generically to refer to both electrons and positrons, if not otherwise stated.}--proton collisions proceeds mainly via photon--photon interactions, $\gamma \gamma \rightarrow \ell^+ \ell^-$, where photons are radiated from the incident beam particles~\cite{Vermaseren:1982cz}. 
The clean experimental signature and the precise SM prediction of this process provide a high sensitivity to searches for new phenomena producing multi-lepton final states.
For example, the production of a doubly charged Higgs boson~\cite{Accomando:1993ar,Aktas:2006nu} or processes involving  generic bosons carrying two units of lepton number (bileptons)~\cite{Cuypers:1996ia}, could lead to multi-lepton events of large invariant mass.
Measurements of both multi-electron~\cite{Aktas:2003jg}  and muon pair~\cite{Aktas:2003sz} production at high transverse momentum ($P_T$) have already been performed by the H1 Collaboration using a data sample corresponding to an integrated luminosity of $115$~pb$^{-1}$.
Events with high invariant mass $M_{12}$ of the two highest $P_T$ electrons were measured. Three events with two electrons and three events with three electrons were observed in the region $M_{12}>100$~GeV, where the SM prediction is low~\cite{Aktas:2003jg}.

In the present paper a measurement of multi-lepton production at high $P_T$ using the complete $e^\pm p$ HERA collider data of the H1 experiment is presented.
The data are recorded at electron and proton beam energies of $27.6$~GeV and $820$~GeV or $920$~GeV, corresponding to centre-of-mass energies $\sqrt{s}$ of $301$~GeV or $319$~GeV, respectively.
The total integrated luminosity of the data is $463$~pb$^{-1}$, which represents a factor of four increase with respect to the previous published results.
The data comprise $178$~pb$^{-1}$ recorded in $e^-p$ collisions and $285$~pb$^{-1}$ in $e^+p$ collisions, of which $35$~pb$^{-1}$ were recorded at $\sqrt{s} = 301$~GeV.
While previous measurements were performed mainly using $e^+p$ data, an $e^-p$ data set with more than a ten-fold increase in integrated luminosity is now analysed.
The current analysis is extended with respect to those presented in~\cite{Aktas:2003jg,Aktas:2003sz}, such that all event final states with at least two high $P_T$ leptons, electrons ($e$) or muons  ($\mu$), are now investigated.
In addition, differential cross sections of the production of $e^+e^-$ and $\mu^+\mu^-$ pairs are measured in a restricted phase space dominated by photon-photon collisions.
%

\section{Standard Model Processes and their Simulation}\label{sec:MC}

Multi-lepton events are simulated with the GRAPE~\cite{Abe:2000cv}
generator, which includes all electroweak matrix elements at tree level.
The production mechanisms include $\gamma \gamma$, $\gamma Z$, $ZZ$
interactions, internal photon conversion and the decay of virtual or
real $Z$ bosons.
Different approaches for the cross section calculation are followed depending on the virtuality $Q^2$ of the photon coupled to the proton and on the mass $W$ of the hadronic final state.
In the elastic region, $ep \rightarrow e \: \ell^+ \ell^- p$, the proton vertex is described in terms of dipole form factors.
The quasi-elastic domain is defined by $W < 5$~GeV or $Q^2 < 1$~GeV$^2$.
In the region $W < 2$~GeV, a resonance parametrisation~\cite{Brasse:1976bf}
is used for the proton vertex.
In the remaining quasi-elastic phase space, a fit to photoproduction and deep inelastic scattering data is used~\cite{Abramowicz:1997ms}.
In the inelastic regime, corresponding to electron-quark interactions with $W > 5$~GeV and $Q^2 > 1$~GeV$^2$, proton parton densities taken from the CTEQ5L~\cite{Pumplin:2002vw} parametrisation are used.
Initial and final state QED radiation is included.  
The hadronic final state is simulated via
interfaces to PYTHIA~\cite{Sjostrand:2000wi} and SOPHIA~\cite{Mucke:1999yb} for the inelastic and quasi-elastic regimes, respectively.  
GRAPE predicts cross sections for
$ep \rightarrow e \: \mu^+ \mu^- X$ and $ep \rightarrow e \: e^+ e^- X$  processes, leading to $e\mu\mu$ and $eee$ final states.
 Events with only two leptons ($\mu\mu$, $e\mu$ or $ee$) are observed if the scattered electron or one lepton of the pair is not detected. 
The $ep \rightarrow e \: \tau^+ \tau^- X$ process with subsequent leptonic tau decays is also simulated with GRAPE and its 
contribution to the studied final states is found to be at most $4$\%.

Experimental backgrounds to $ee$ and $eee$ final states arise from events in which, in addition to a genuine electron, one or two fake electrons are reconstructed.
Neutral current (NC) deep-inelastic scattering (DIS) events ($ep \rightarrow e X$) in which hadrons or radiated photons are wrongly identified as electrons constitute the dominant background contribution.
QED Compton scattering $ep \rightarrow e \gamma X$ may also contribute if the photon is misidentified as an electron. 
Background to the $e\mu$ final state may arise from NC DIS events if hadrons are misidentified as muons.
The NC DIS and Compton processes are
simulated using the RAPGAP~\cite{Jung:1993gf} and WABGEN~\cite{Berger:kp}
Monte Carlo (MC) generators, respectively.

Generated events are passed through the full GEANT~\cite{Brun:1987ma} based simulation of the H1 apparatus, which takes into account the running conditions of the different data taking periods, and are reconstructed and analysed with the same program chain as is used for the data.

\section{Experimental Conditions}

A detailed description of the H1 experiment can be found in~\cite{Abt:h1}.
Only the detector components relevant to the
present analysis are briefly described here.  
The origin of the H1 coordinate system is the nominal $ep$ interaction point, with the direction of the proton beam defining the positive $z$-axis (forward region). Transverse momentum is measured in the $xy$ plane. The pseudorapidity $\eta$ is related to the polar angle $\theta$ by $\eta = -\ln \, \tan (\theta/2)$.
The Liquid Argon (LAr) calorimeter~\cite{Andrieu:1993kh} covers the polar angle range
$4^\circ < \theta < 154^\circ$ with full azimuthal acceptance.
Electromagnetic shower energies are measured with a precision of
$\sigma (E)/E = 12\%/ \sqrt{E/\mbox{GeV}} \oplus 1\%$ and hadronic energies
with $\sigma (E)/E = 50\%/\sqrt{E/\mbox{GeV}} \oplus 2\%$, as measured in test beams~\cite{Andrieu:1993tz,Andrieu:1994yn}.
In the backward region, energy measurements are provided by a lead/scintillating-fiber (SpaCal) calorimeter~\cite{Appuhn:1996na} covering the range $155^\circ < \theta < 178^\circ$.
The central ($20^\circ < \theta < 160^\circ$)  and forward ($7^\circ < \theta < 25^\circ$)  inner tracking detectors are used to
measure charged particle trajectories, to reconstruct the interaction
vertex and to complement the measurement of hadronic energies.
The LAr calorimeter and inner tracking detectors are enclosed in a super-conducting magnetic
coil with a field strength of $1.16$~T.
From the curvature of charged particle trajectories in the magnetic field, the central tracking system provides transverse momentum measurements with a resolution of $\sigma_{P_T}/P_T = 0.005 P_T / \rm{GeV} \oplus 0.015$~\cite{Kleinwort:2006zz}.
The return yoke of the magnetic coil is the outermost part of the detector and is
equipped with streamer tubes forming the central muon detector
($4^\circ < \theta < 171^\circ$).

The luminosity is determined from the rate of the Bethe-Heitler process $ep \rightarrow ep \gamma$,
measured using a photon detector located close to the beam pipe at $z=-103~{\rm m}$, in the backward direction.

Events having an electromagnetic deposit in the LAr with an energy greater than $10$~GeV are detected by the LAr trigger with an efficiency close to $100$\%~\cite{Adloff:2003uh}.
The muon trigger is based on single muon signatures from the central muon detector, combined with signals from the central tracking detector. The trigger efficiency is about $90$\% for di-muon events with muon transverse momenta larger than $10$ and $5$~GeV.
A combination of the LAr and muon triggers is used if an electron and a muon are both present in the final state, resulting in an efficiency close to $100$\% for electron and muon transverse momenta above $5$~GeV.

\section{Data Analysis}

\subsection{Particle Identification}\label{sec:part_def}

Electron candidates are identified in the polar angle range $5^\circ < \theta < 175^\circ$ as a compact and isolated electromagnetic shower in either the LAr or SpaCal calorimeter.  
The electron energy and angular direction are measured by the calorimeters.
The electron energy threshold is $5$~GeV in the polar angle range $20^\circ < \theta < 175^\circ$ and is raised to $10$~GeV in the forward region ($5^\circ < \theta < 20^\circ$).
The calorimetric energy measured within a distance in the pseudorapidity-azimuth $(\eta - \phi)$ plane $R=\sqrt{\Delta \eta^2 + \Delta \phi^2} < 0.75$ around the electron is required to be below $2.5$\% of the electron energy.
In the region of angular overlap between the LAr and
the central tracking detectors ($20^\circ < \theta < 150^\circ$), hereafter referred to as the central region, the
calorimetric electron identification is complemented by tracking information.  
In this region it is required that a well measured track
geometrically matches the centre-of-gravity of the electromagnetic cluster within a distance of
closest approach of $12$~cm.  
The electron is required to be isolated from any other well measured track by a distance $R > 0.5$ to the electron direction.
Furthermore,
the distance from the first measured track point in the central drift
chambers to the beam axis is required to be below $30$~cm in order to
reject photons that convert late in the central tracker material. 
In addition, in the central region the transverse momentum of the associated electron track $P_T^{e_{tk}}$ is required to match the
calorimetric measurement $P_T^e$ such that $1/P_T^{e_{tk}} - 1/P_T^e <
0.02$~GeV$^{-1}$ in order to reject hadronic showers.  
Due to a lower track reconstruction efficiency and higher showering probability, no track conditions are required for electron candidates in the forward ($5^\circ < \theta < 20^\circ$) and backward ($150^\circ <\theta < 175^\circ$) regions. 
In these regions, no distinction between electrons and photons is made.
The resulting electron identification efficiency is $80$\% in the central region and larger than $95$\% in the forward and backward regions.

Muon candidates are identified in the polar angle range $20^\circ <\theta < 160^\circ$ with a minimum transverse momentum of $2$~GeV.
Muon identification is based on the measurement of a track in the inner tracking systems associated with a track segment or an energy deposit in the central muon detector~\cite{Aktas:2003sz,Andreev:2003pm}.
The muon momentum is measured from the inner track curvature in the solenoidal magnetic field.
A muon candidate should have no more than $3.5$~GeV deposited in a
cylinder, centred on the muon track direction, of radius $25$~cm and $50$~cm in the electromagnetic and hadronic sections of the LAr calorimeter, respectively.
Misidentified hadrons are strongly suppressed by requiring that the muon candidate be separated from any jet and from any track by $R > 1$.
The efficiency to identify muons is $\sim 90$\%.

Calorimeter energy deposits and tracks not previously identified as electron or muon candidates are used to form combined cluster-track objects, from which the hadronic energy is reconstructed~\cite{matti,benji}.
Jets are reconstructed from these combined cluster-track objects using an inclusive $k_T$ algorithm~\cite{Ellis:1993tq,Catani:1993hr} with a minimum transverse momentum of $2.5$~GeV.

\subsection{Event Selection}

In order to remove background events induced by cosmic rays and other non-$ep$ sources, the event vertex is required to be reconstructed within $35$~cm in $z$ of the nominal interaction point. In addition, to remove non-$ep$ background topological filters and timing vetoes are applied.

Multi-lepton events are selected by requiring at least two
central ($20^\circ < \theta < 150^\circ$) electron or muon
candidates, of which one must have $P_T^\ell > 10$~GeV and the other
$P_T^{\ell} > 5$~GeV. 
Additional lepton candidates are identified in the
detector according to the criteria defined in section~\ref{sec:part_def}.
All lepton candidates are required to be isolated with respect to each other by a minimum distance in pseudorapidity-azimuth of $R > 0.5$.
Lepton candidates are
ordered according to decreasing transverse momentum, $P_T^{\ell_i} > P_T^{\ell_{i+1}}$.
Final states with all possible combinations of lepton candidates are investigated.
Selected events are classified into independent samples according to the flavour and the number of lepton candidates (e.g. $ee$, $e\mu$, $e\mu\mu$).

In order to measure the lepton pair production cross section in a well
defined region of phase space, sub-samples of $ee$ and $\mu\mu$ events dominated by
photon-photon collisions are selected, labelled $(\gamma\gamma)_{e}$ and $(\gamma\gamma)_{\mu}$, respectively.
In these subsamples the two leptons are required to be of opposite charge and a significant deficit compared to the initial state must be observed in the difference\footnote{For fully contained events or events where only longitudinal momentum along the proton direction ($+ z$) is undetected, one expects $E-P_z = 2E^0_e = 55.2$~GeV, where $E^0_e$ is the energy of the incident electron. If the scattered electron is undetected, the threshold $E-P_z < 45$~GeV corresponds to a cut on the fractional energy loss $y=(E-P_z)/2E^0_e < 0.82$.} $E -P_z$ of the energy and the longitudinal momentum of all visible particles, $E -P_z <45$~GeV.
These two conditions ensure that the incident electron is lost in the beam pipe after radiating a quasi-real photon of squared four-momentum $Q^2$ lower than $1$~GeV$^2$.

\subsection{Systematic Uncertainties}
\label{sec:syst}

The following experimental systematic uncertainties are considered:

\begin{itemize}
\item The uncertainty on the electromagnetic energy scale varies depending on the polar angle from $0.7$\% in the backward and central region to $2$\% in the forward region. The polar angle measurement  uncertainty of electromagnetic clusters is $3$~mrad. The identification efficiency of electrons is known with an uncertainty of $3$ to $5$\%, depending on the polar angle. 
\item The scale uncertainty on the transverse momentum of high $P_T$ muons is $2.5$\%. The uncertainty on the reconstruction of the muon polar angle is $3$~mrad. The identification efficiency of muons is known with an uncertainty of $5$\%.
\item The hadronic energy scale is known within $2$\% at high transverse momentum and $5$\% for events with a total hadronic transverse momentum below $8$~GeV. 
\item The uncertainty on the trigger efficiency is estimated to be $3$\% if at least one electron candidate is detected, and $6$\% if only muons are present in the final state.
\item The luminosity measurement has an uncertainty of $3$\%.
\end{itemize}

The effect of the above systematic uncertainties on the SM expectation is determined by varying the experimental quantities by $\pm 1$ standard deviation in the MC samples and propagating these variations through the whole analysis chain.

Additional model systematic uncertainties are attributed to the SM Monte Carlo generators described in section~$\ref{sec:MC}$.
The theoretical uncertainty on the lepton pair production cross section calculated with GRAPE is $3$\%.
The uncertainty on the QED Compton and NC DIS background is $20$\%, as deduced from dedicated studies~\cite{Aktas:2003jg}.
The total error on the SM prediction is determined by adding the effects of all model and experimental systematic uncertainties in quadrature.

\section{Results}

\subsection{Multi--Lepton Event Samples}

\begin{figure}[!htbp] 
  \begin{center}
  \begin{center}
 \includegraphics[width=.5\textwidth]{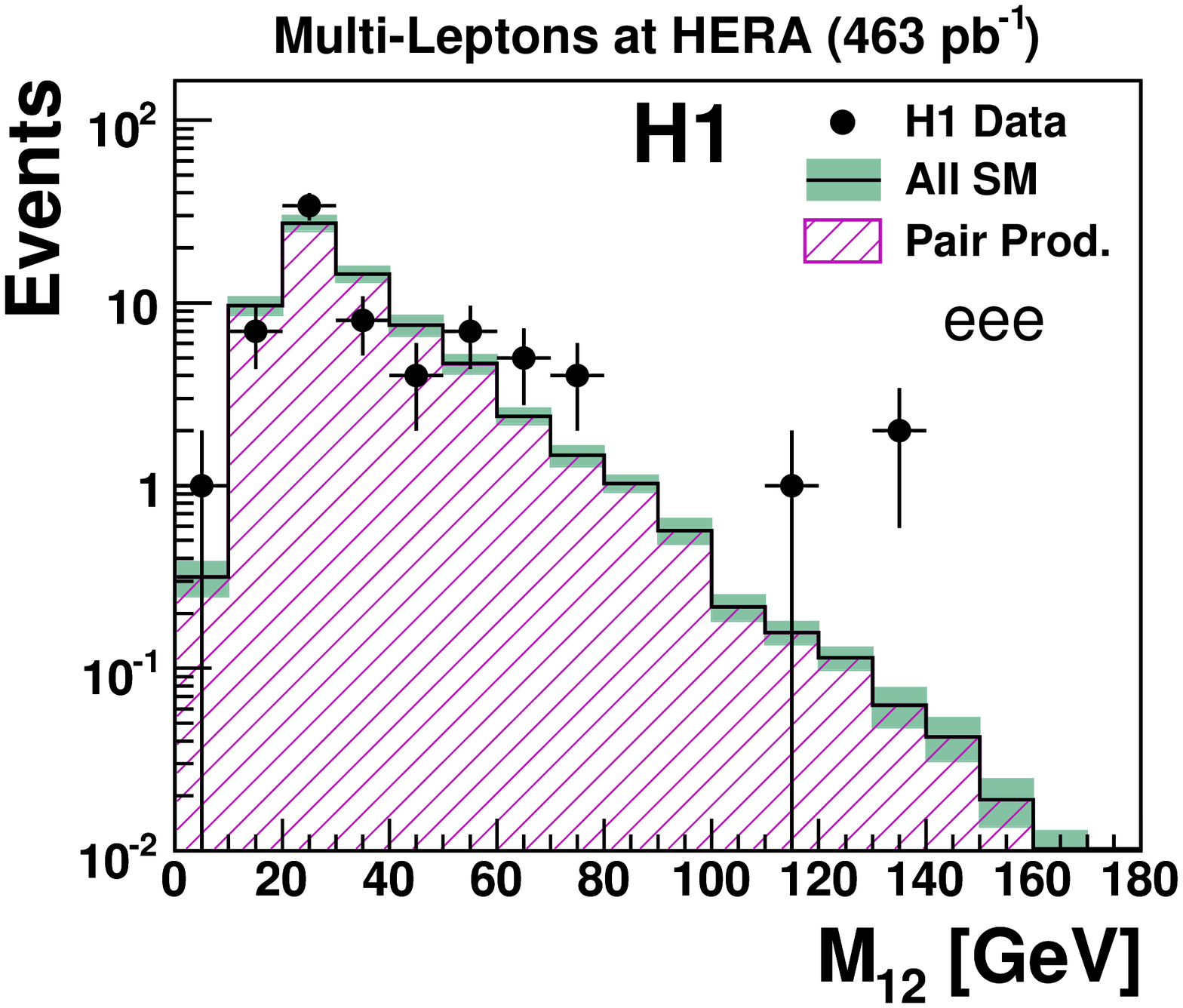}\put(-11,33){{(a)}}
 \includegraphics[width=.5\textwidth]{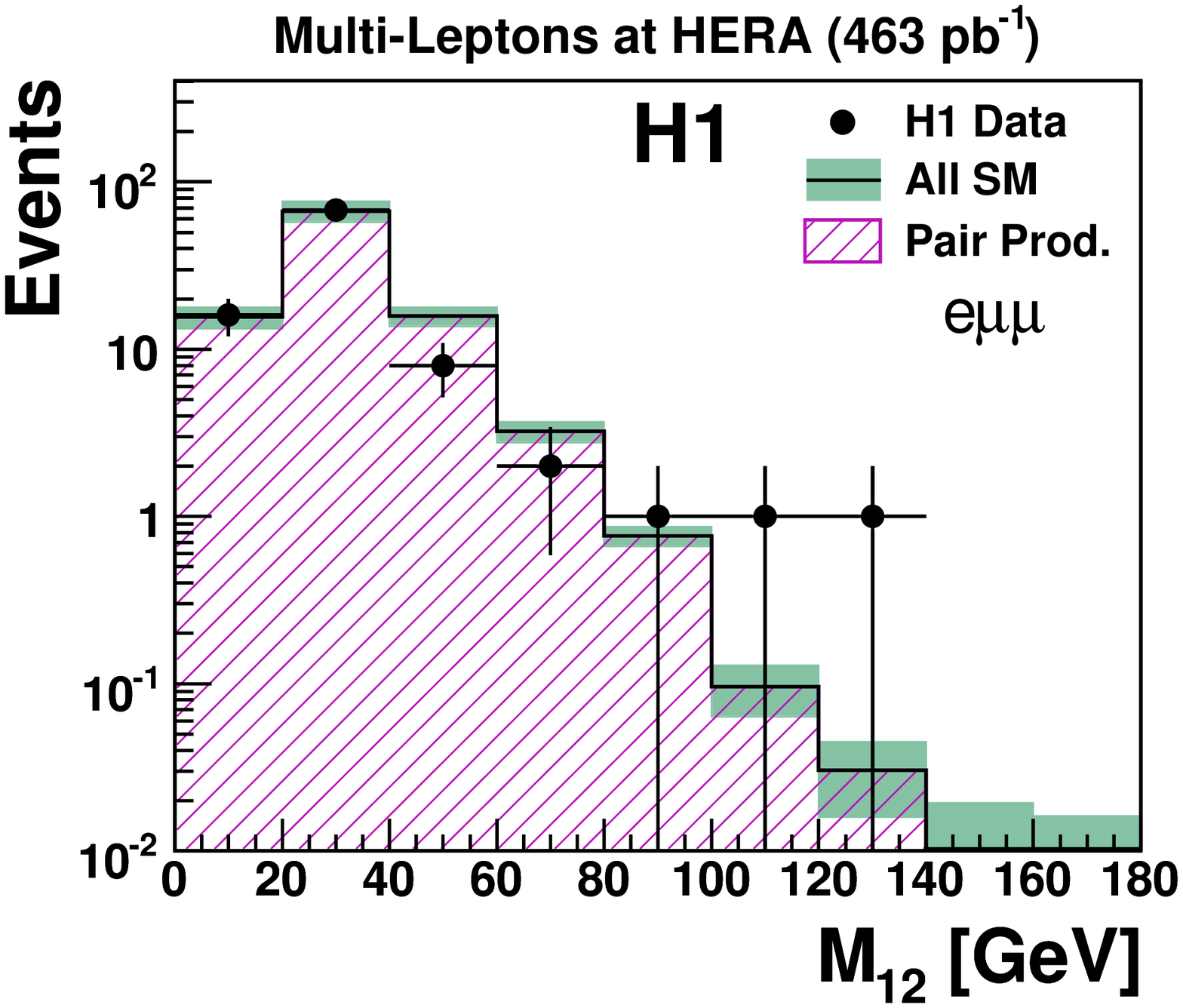}\put(-11,33){{(b)}}\\
 \includegraphics[width=.5\textwidth]{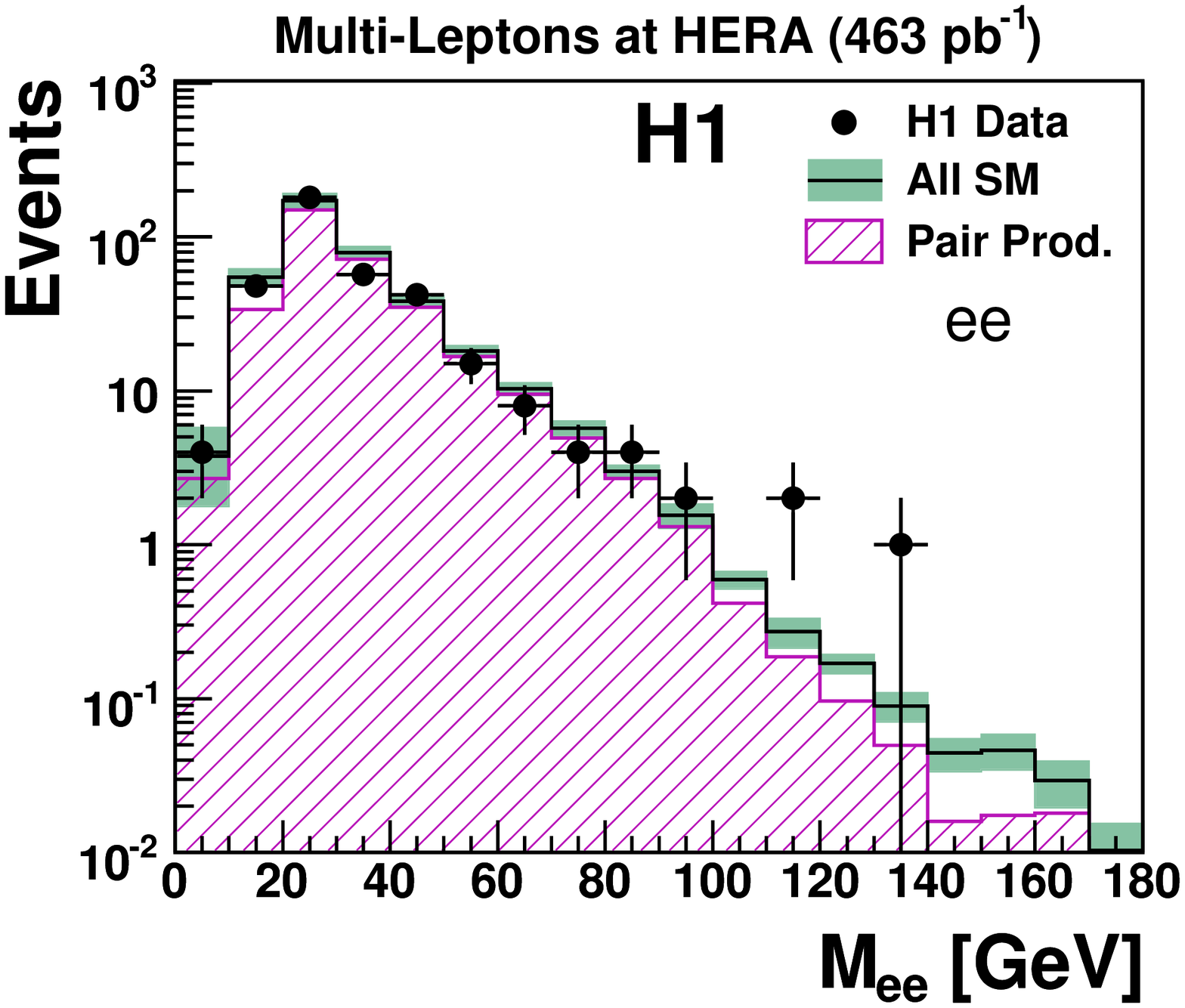}\put(-11,33){{(c)}}
 \includegraphics[width=.5\textwidth]{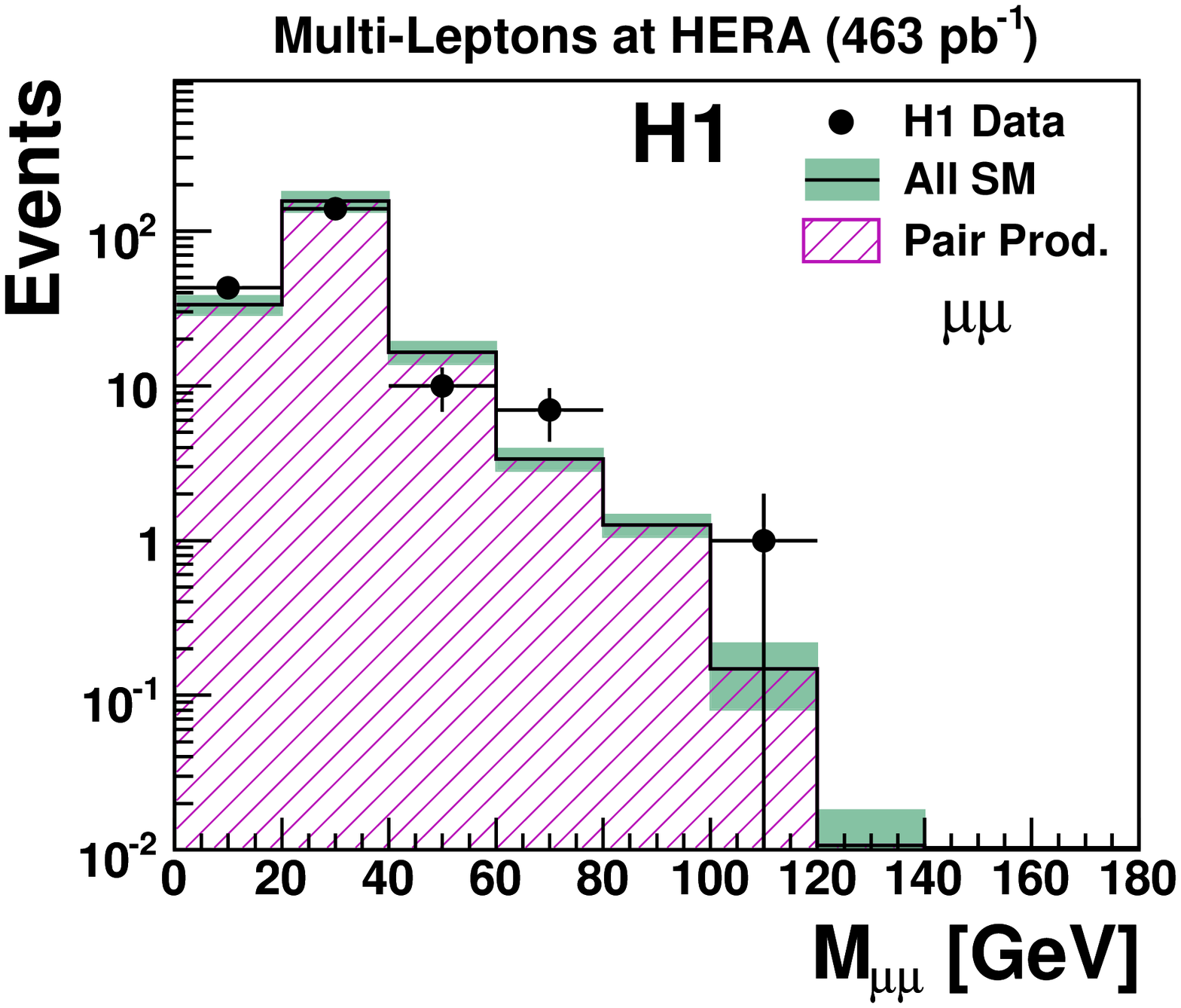}\put(-11,33){{(d)}}\\
\hspace{-.5\textwidth} \includegraphics[totalheight=6.4cm]{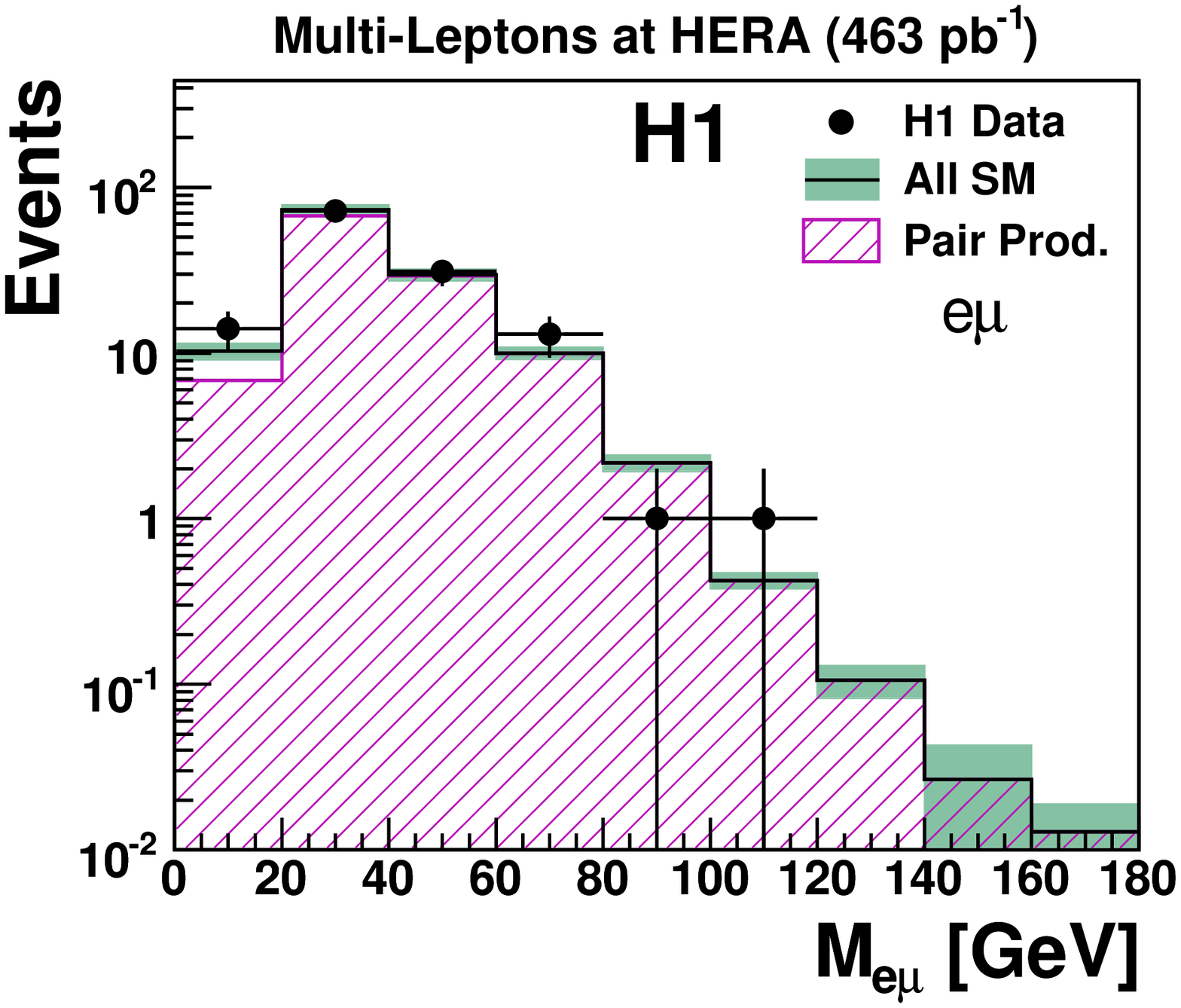}\put(-11,33){{(e)}}
 \end{center}
  \end{center}
  \caption{The distribution of the invariant mass of the two highest $P_T$ leptons for events classified as $eee$ (a), $e\mu\mu$ (b) and $ee$ (c), $\mu\mu$ (d) and $e\mu$ (e). 
  The points correspond to the observed data events and the open histogram to the SM expectation. The total error on the SM expectation is given by the shaded band. The component of the SM expectation arising from lepton pair production is given by the hatched histogram.
   }
\label{fig:Masses}  
\end{figure}

The observed event yields are summarised in table~\ref{tab:mlepyields}.
Only classes for which at least one data event is selected are shown.
In all other classes, no event is observed and the SM prediction is also negligible.
The observed numbers of events are in good agreement with the SM expectations.
The $eee$ and $ee$ channels are dominated by electron pair production.
The $e\mu\mu$, $\mu\mu$ and $e\mu$ channels contain mainly events from muon pair production.
The $e\mu$ channel is populated if the scattered electron and only one muon of the pair is selected.
Four data events are classified as $ee\mu$ compared to a SM expectation of $1.43 \pm 0.26$, dominated by the production of muon pair events where one muon is lost and the second electron candidate is due to a radiated photon. 
One event with four electron candidates is observed compared to a prediction of $0.33 \pm 0.07$. %
According to MC simulations, this signature is due to 
tri-electron events with a radiated photon.
The distributions of the invariant mass $M_{12}$ of the two highest
$P_T$ leptons for the $eee$ and $e\mu\mu$ samples are shown in figures~\ref{fig:Masses}(a) and (b), respectively.
The distributions of the invariant mass $M_{12}$  of the two leptons in the
di-lepton event classes are presented in figures~\ref{fig:Masses}(c), (d) and (e). 
An overall agreement with the SM prediction is observed in all cases.

\begin{figure}[!htbp] 
  \begin{center}
 \includegraphics[width=.65\textwidth]{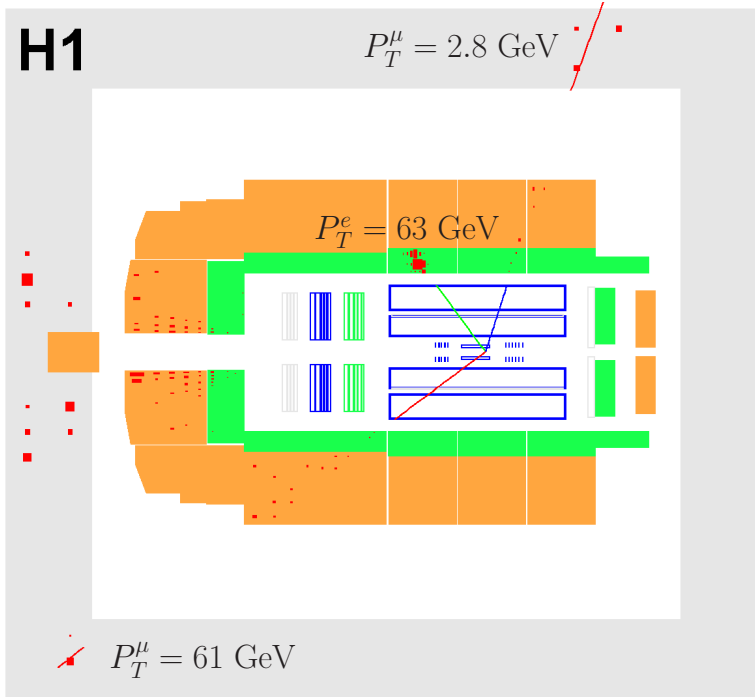}
  \end{center}
  \caption{Display (side view) of the $e\mu\mu$ event observed with the highest $M_{12}$ invariant mass. Indicated are the reconstructed tracks and the energy depositions in the calorimeters. The beam positrons enter the detector from the left and the protons from the right.
   }
\label{fig:display}  
\end{figure}

High invariant mass events ($M_{12} > 100$~GeV) are observed in the data. 
The corresponding observed and predicted event yields are summarised for all channels in table~\ref{tab:mlepyieldsM100}.
The three $ee$, three $eee$ and one $\mu\mu$ high mass events have already been discussed extensively in previous H1 publications~\cite{Aktas:2003jg,Aktas:2003sz}. 
No additional events in these classes are observed in the new data.
One $e\mu$ and two $e\mu\mu$ high mass events are observed in the new data.
The $e\mu\mu$ event with the largest invariant mass $M_{12} = 127 \pm 10$~GeV is presented in figure~\ref{fig:display}.
In this event, $M_{12}$ is formed by the electron and the highest $P_T$ muon.
In the other $e\mu\mu$ event, $M_{12}$ is formed by the two muons. 
No $eeee$ or $ee\mu$ event is observed with a di-lepton invariant mass above $100$~GeV, in agreement with the corresponding SM expectations below $0.01$.

The results for $e^+p$ and $e^-p$ data are shown separately in table~\ref{tab:mlepyieldsM100}. 
The di-lepton ($ee$, $e\mu$ and $\mu\mu$) and tri-lepton ($eee$ and $e\mu\mu$) events  at high mass $M_{12}>100$~GeV are all observed in $e^+p$ collisions 
whereas no such event is observed in  the $e^-p$ data.

\begin{figure}[!htbp] 
  \begin{center}
\includegraphics[width=.5\textwidth]{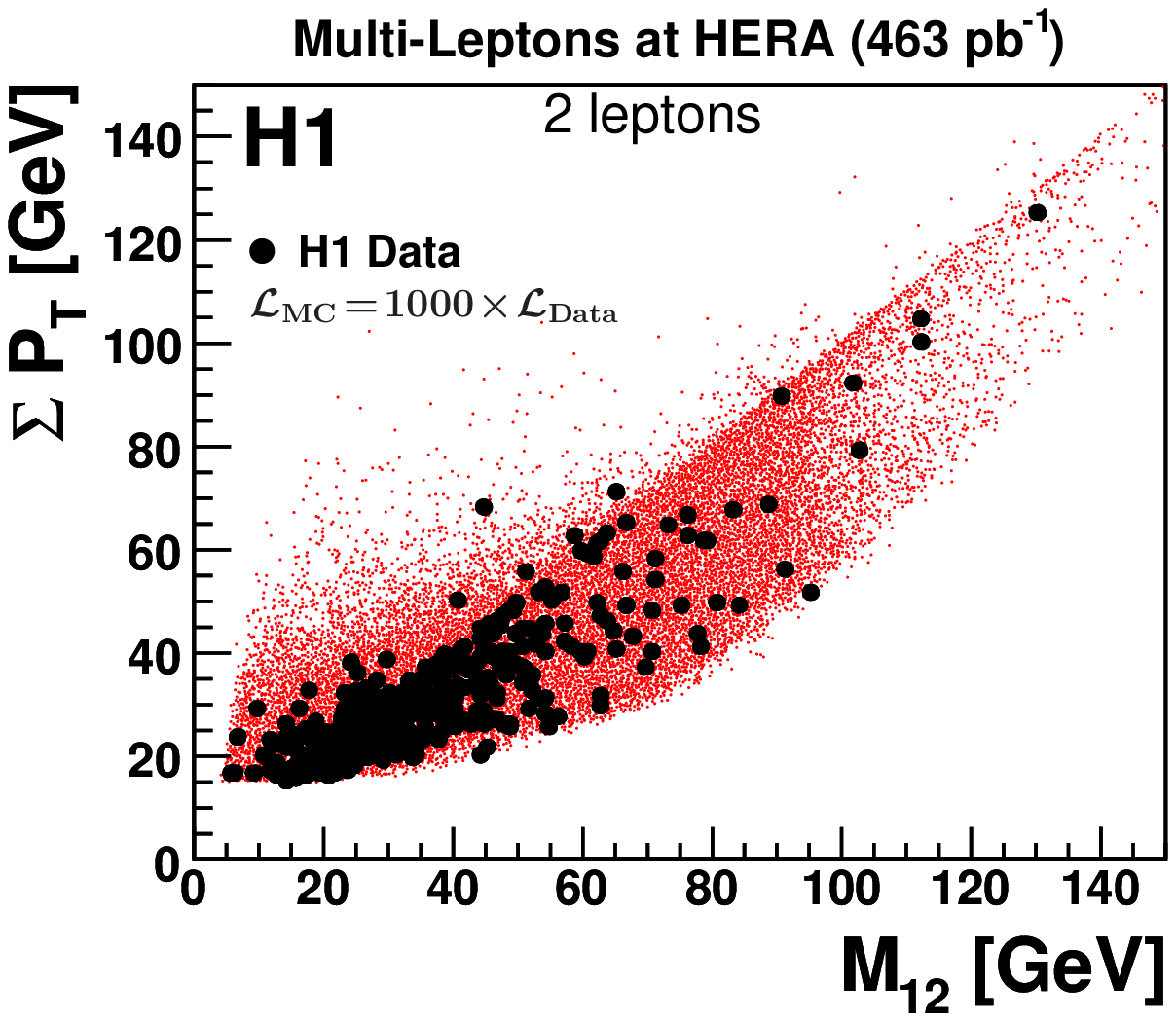}\put(-11,15){{(a)}}
 \includegraphics[width=.5\textwidth]{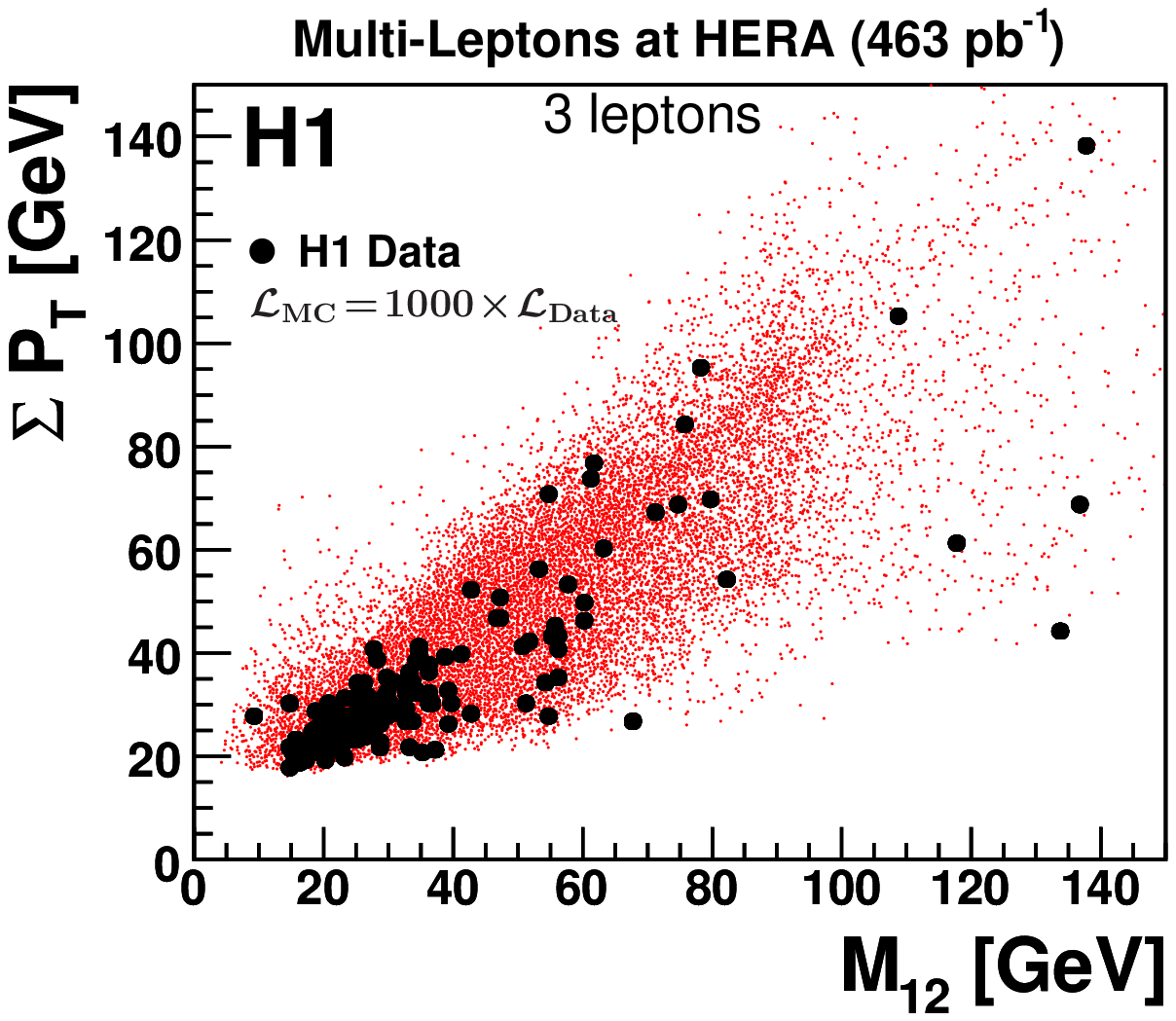}\put(-11,15){{(b)}}\\
  \end{center}
  \caption{Correlation of the invariant mass $M_{12}$ with the scalar sum of the transverse momenta $\sum P_T$ for di-lepton (a) and tri-lepton (b) events. The bold dots represent the data while the small points represent the pair production (GRAPE) prediction for a luminosity $\sim 1000$ times higher than that of data.
   }
\label{fig:2Dcorrelations}  
\end{figure}

The topology of the multi-lepton events can be further investigated using 
the scalar sum of the lepton transverse momenta $\sum P_T$. 
This variable indicates the ``hardness'' of the event and also offers a good sensitivity for new physics searches~\cite{Aktas:2004pz}. 
Figure~\ref{fig:2Dcorrelations} presents the correlation between $\sum P_T$ and the invariant mass $M_{12}$, separately for di-lepton and tri-lepton events.
It can be seen that tri-lepton events may have a large $M_{12}$ but only intermediate $\sum P_T$. 
In such cases, the high mass is formed by one forward and one backward lepton.  
However, for di-lepton event classes high $M_{12}$ values also imply  a large $\sum P_T$.

\begin{figure}[htbp] 
\begin{center}
 \includegraphics[width=.5\textwidth]{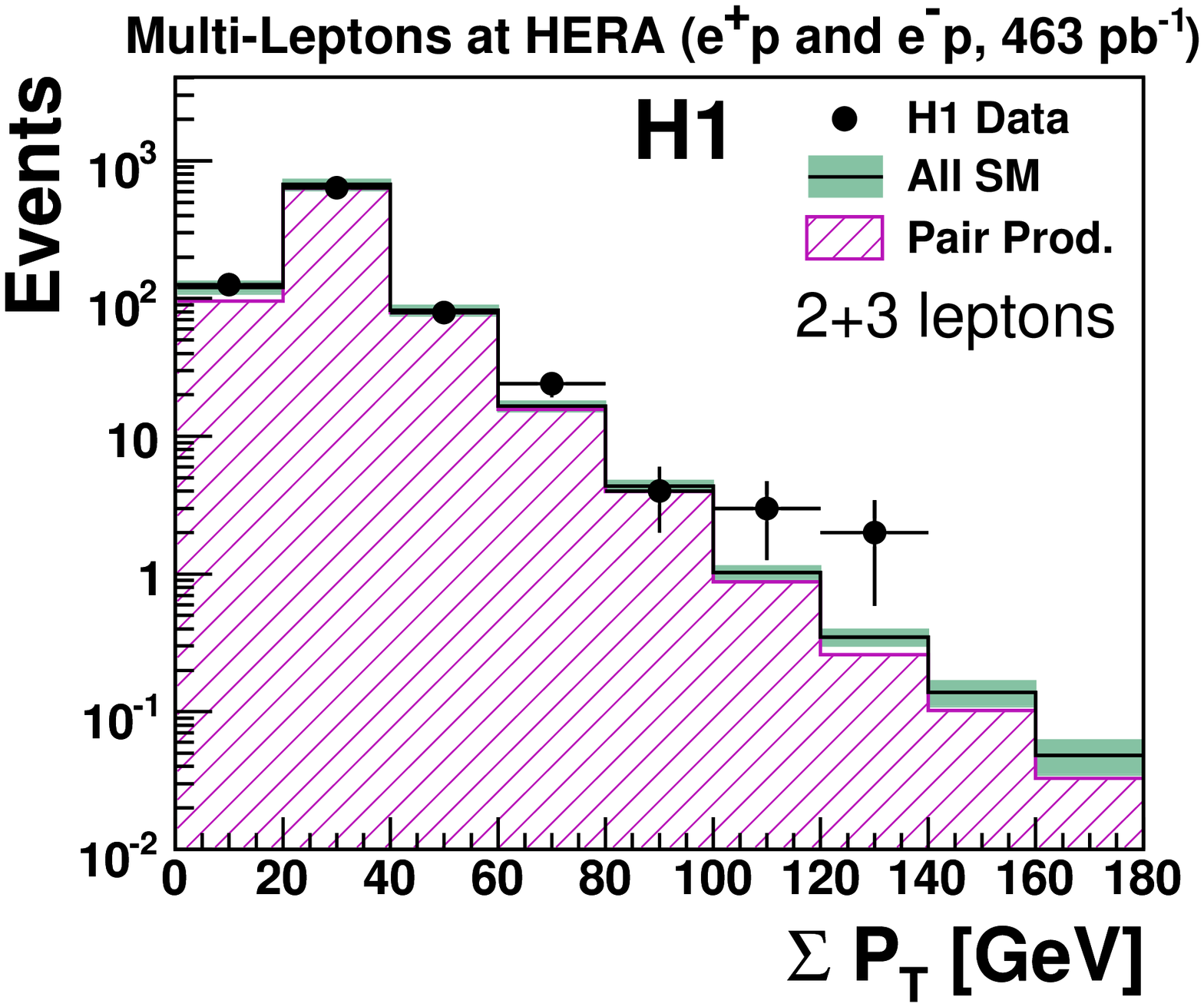}\put(-11,33){{(a)}}\\
 \includegraphics[width=.5\textwidth]{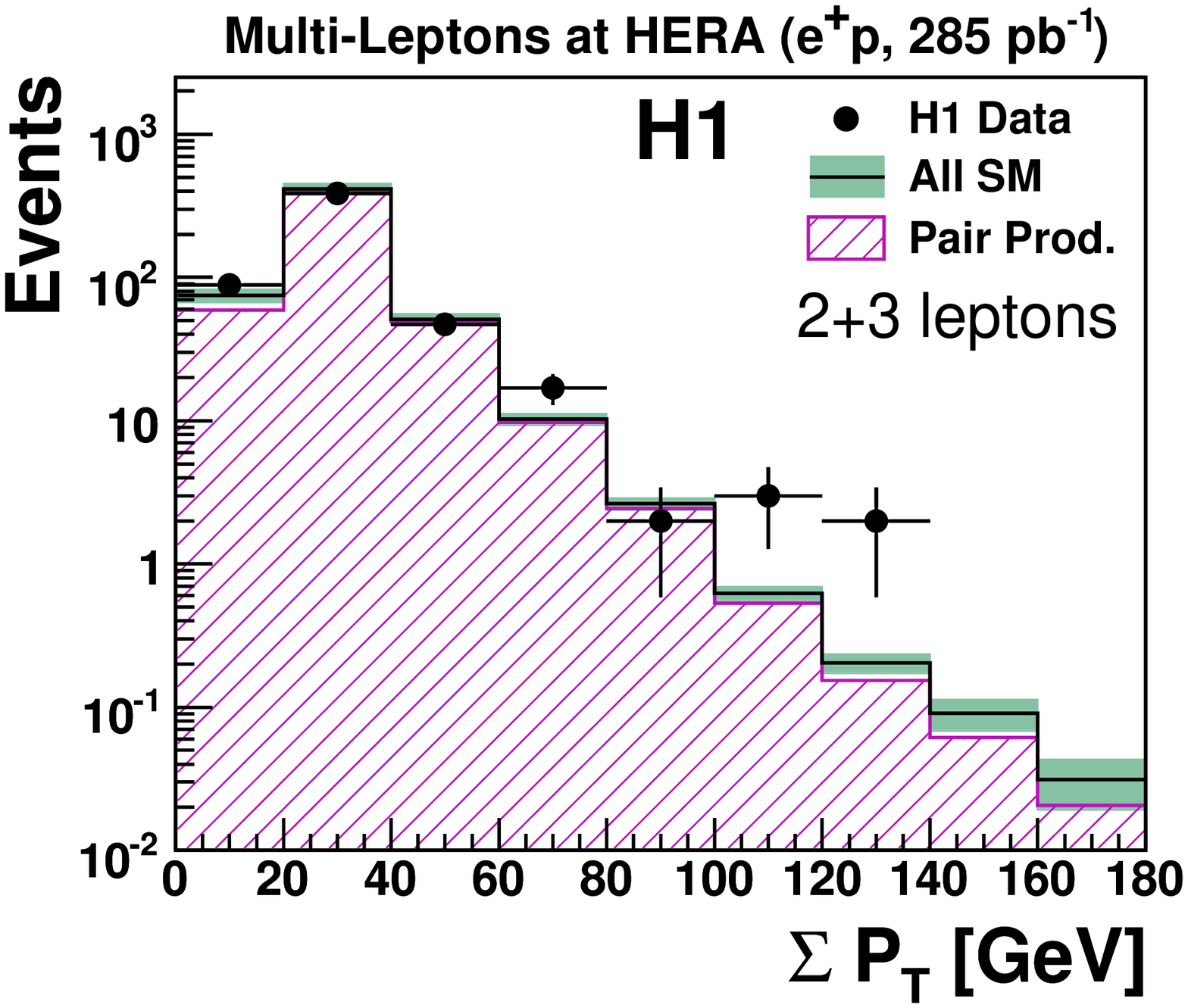}\put(-11,33){{(b)}}\\
 \includegraphics[width=.5\textwidth]{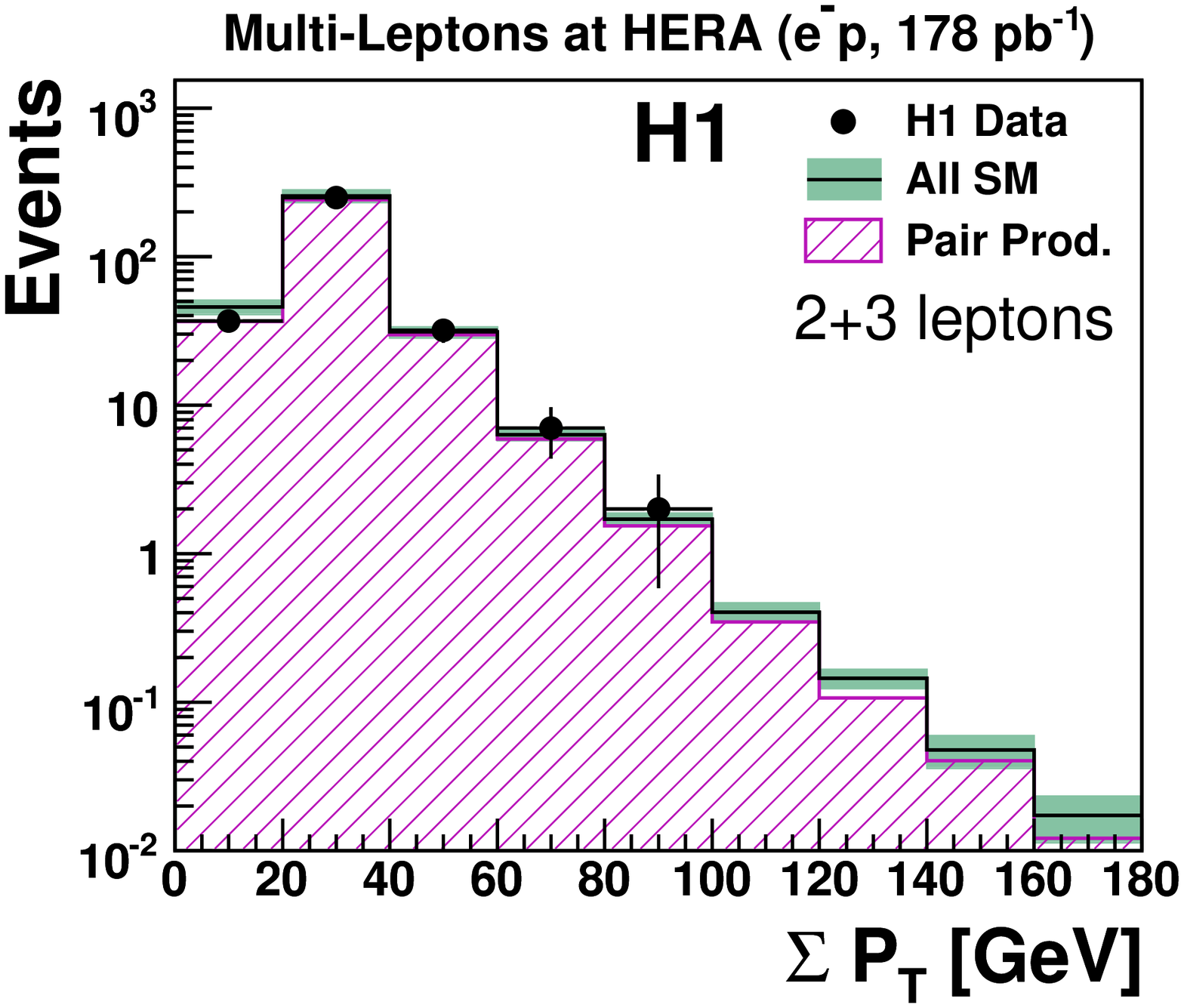}\put(-11,33){{(c)}}
\end{center}
\vspace{-0.5cm}
  \caption{The distribution of the scalar sum of the transverse momenta $\sum P_T$ for combined di-lepton and tri-lepton event samples for all data (a) as well as for $e^+p$ (b) and $e^-p$ (c).
  The points correspond to the observed data events and the open histogram to the SM expectation. The total error on the SM expectation is given by the shaded band. The component of the SM expectation arising from lepton pair production is given by the hatched histogram.
}
\label{fig:SumEt_All_lep}
\end{figure}

The quantity $\sum P_T$ can be used to select the most energetic events and allows the combination of different topologies  of di-lepton and tri-lepton events with electrons and muons.
Figure~\ref{fig:SumEt_All_lep} presents the distributions of $\sum P_T$ of the observed multi-lepton events compared to the SM expectation. 
A good overall agreement between the data and the SM prediction is observed.
For $\sum P_T >$~$100$~GeV, five events are
observed in total, compared to $1.60 \pm 0.20$ expected from the SM (see table~\ref{tab:mlepyieldsEt100}). 
These five events were all recorded in the $e^+p$ data, for which the SM expectation is $0.96$~$\pm$~$0.12$. 
Furthermore, the events correspond to the three $ee$ and the two $e\mu\mu$ events observed with $M_{12} > 100$~GeV.

\subsection{Cross Section Measurements}

Cross sections of the production of electron and muon pairs from photon-photon collisions are measured using the selected $(\gamma\gamma)_{e}$ and $(\gamma\gamma)_{\mu}$ samples.
The kinematic domain of the measurement is defined by $20^\circ < \theta^{\ell_{1,2}} < 150^\circ$, $P_T^{\ell_1} > 10$~GeV,  $P_T^{\ell_2} > 5$~GeV, $Q^2 < 1$~GeV$^2$ and $y < 0.82$.
The data samples collected at $\sqrt{s} = 301$~GeV and $319$~GeV are combined taking into account their respective luminosities.
Assuming a linear dependence of the cross section on the proton beam energy, as predicted by the SM, the resulting cross section corresponds to an effective $\sqrt{s} = 318$~GeV.
The total numbers of observed  $(\gamma\gamma)_{e}$ and $(\gamma\gamma)_{\mu}$ events are in agreement with the SM expectations, as summarised in table~\ref{tab:mlepyields}. 
In the $(\gamma\gamma)_{e}$ sample, the contamination from NC DIS and QED Compton background events is  $2$\%.
No significant background is present in the $(\gamma\gamma)_{\mu}$ sample.

The cross section is evaluated in each bin $i$ using the formula

\begin{equation}
\sigma_i = \frac{N_i^{\rm{data}}-N_i^{\rm{bgr}}}{ {\cal L} \cdot A_i},
\label{eq:xsection}
\end{equation}

\noindent where $N_i^{\rm{data}}$ is the number of observed events in bin $i$, $N_i^{\rm{bgr}}$ the expected contribution from background processes in bin $i$, ${\cal L}$ the integrated luminosity of the data and $A_i$ the signal acceptance in bin $i$. 
The signal acceptance is calculated using GRAPE MC events, as the ratio of the number of events reconstructed in bin $i$ divided by the number of events generated in the same bin. 
It accounts for detection efficiencies and migrations between bins.
The mean signal acceptance is $45$\% for $ep \rightarrow e \: e^+e^- X$ events and $60$\% for $ep \rightarrow e \: \mu^+\mu^- X$ events.
The systematic error of the measured cross section is determined by repeating the analysis after applying the appropriate variations to the MC for each systematic source, as described in section~\ref{sec:syst}.

The measured $ep \rightarrow e \: e^+e^- X$ cross section integrated over the phase space defined above is $\sigma = 0.67 \pm 0.06 \pm 0.05$~pb, where the first error is statistical and the second systematic. 
The measured cross section for muon pair production, $ep \rightarrow e \: \mu^+\mu^- X$, in the same phase space is $\sigma = 0.63 \pm 0.05 \pm 0.09$~pb.
The results are in agreement with the SM expectation of $0.63 \pm 0.02$~pb calculated using the GRAPE generator.
Combining the di-electron and di-muon samples, an average lepton pair production cross section of $\sigma = 0.65 \pm 0.04 \pm 0.06$~pb is measured.

\begin{figure}[htbp] 
\begin{center}
\includegraphics[totalheight=13cm]{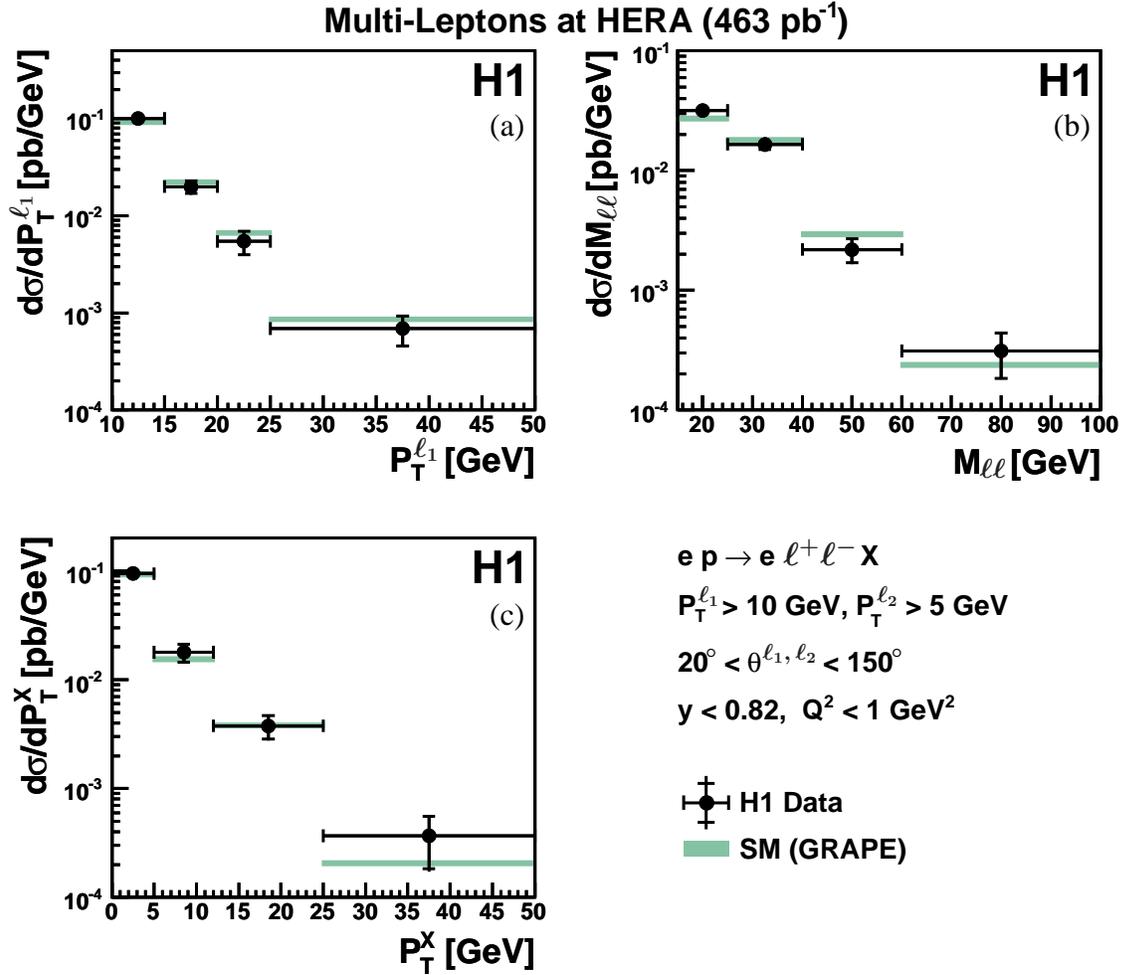}\put(-87,47){{(c)}}\put(-87,112){{(a)}}\put(-12,112){{(b)}}
\end{center}
  \caption{The measured cross section for lepton pair production in a restricted phase space dominated by the photon-photon process as a function of the leading lepton transverse momentum $P_T^{\ell_1}$ (a), the invariant mass of the lepton pair $M_{\ell\ell}$ (b) and the hadronic transverse momentum $P_T^X$ (c).
The differential cross section is averaged over the intervals shown.
The inner error bars represent the statistical errors, the outer error bars the statistical and systematic errors added in quadrature.
The bands represent the SM prediction with its one standard deviation uncertainty.  }
\label{fig:XSec_comb}
\end{figure}

The differential cross sections of lepton pair production as a function of the transverse momentum of the leading lepton $P_T^{\ell_1}$, the invariant mass of the lepton pair $M_{\ell\ell}$ and the hadronic transverse momentum $P_T^X$ are listed for each sample in table~\ref{tab:xsection}  and shown in figure~\ref{fig:XSec_comb} for the combined electron and muon samples.
The measurements are in good agreement with the SM cross sections.

\section{Conclusion}

The production of multi-lepton (electron or muon) events at high transverse momenta is studied in $e^+p$ and $e^-p$ scattering.
The full $e^\pm p$ data sample collected by the H1 experiment at HERA with an integrated luminosity of $463$~pb$^{-1}$ is analysed.
The yields of di-lepton and tri-lepton events are in good agreement with the SM predictions.
In each sample distributions of the invariant mass $M_{12}$ of the two highest $P_T$ leptons and of the scalar sum of the lepton transverse momenta $\sum P_T$ are studied and found to be in good overall agreement with the SM expectation. 

Events are observed in the di-lepton and tri-lepton channels with high invariant masses $M_{12}$ above $100$~GeV. All such events are observed in $e^+p$ collisions.
Five of them have a $\sum P_T > 100$~GeV, whereas the corresponding SM expectation for $e^+p$ collisions is $0.96 \pm 0.12$. 
Differential cross sections for electron and muon pair production are measured in a restricted phase space dominated by photon-photon interactions.
The measured cross sections are in agreement with the SM expectations. 
%

\section*{Acknowledgements}

We are grateful to the HERA machine group whose outstanding
efforts have made this experiment possible. 
We thank the engineers and technicians for their work in constructing 
and maintaining the H1 detector, our funding agencies for financial 
support, the DESY technical staff for continual assistance and the 
DESY directorate for the hospitality which they extend to the non DESY 
members of the collaboration.


\clearpage

\begin{table}[]
\begin{center}
\begin{tabular}{ c c c c c }
\multicolumn{5}{c}{Multi-Leptons at HERA ($463$ pb$^{-1}$)}\\
\hline
Selection & Data & SM & Pair Production (GRAPE) & NC DIS + Compton \\
\hline                                        
$ee$ & $368$ & $390 \pm 46$ & $332 \pm 26$  & $58 \pm 30$ \\ 
$\mu\mu$ & $201$   & $211 \pm 32$ & $211 \pm 32$ &  $< 0.005$ \\
$e\mu$ & $132$   & $128 \pm 9$~~ & $118 \pm 8$~~ & $\!\!\!10.0 \pm 2.5$ \\
%
$eee$ & $73$ & $70 \pm 7$ & $69.8 \pm 7.0$ & $0.2 \pm 0.1$ \\    
$e\mu\mu$ & $97$ & $102 \pm 14$  & $102 \pm 14$  &  $< 0.005$    \\   
$ee\mu$   & $4$  & ~~$1.43 \pm 0.26$ & ~~$1.18 \pm 0.20$ & $0.25 \pm 0.14$ \\ 
$eeee$   &  $1$  & ~~$0.33 \pm 0.07$ & ~~$0.33 \pm 0.07$ & $< 0.005$  \\ 
 
\hline
$(\gamma\gamma)_{e}$ & $146$  & $138 \pm 12$ & $135 \pm 11$ & $3.0 \pm 1.0$\\ 
$(\gamma\gamma)_{\mu}$ & $163$ & $162 \pm 24$ & $162 \pm 24$ & $< 0.005$  \\  
\hline
\end{tabular}
\end{center}
\caption{Observed and predicted event yields for the different event classes.
  The errors on the predictions include model uncertainties and experimental systematic errors added in quadrature. The limits on the background estimations correspond to the selection of no event in the simulated sample and are quoted at $95$\% confidence level.}
\label{tab:mlepyields}
\end{table}

\begin{table}[]
\begin{center}
\begin{tabular}{ c c c c c }
\multicolumn{5}{c}{Multi-Leptons at HERA ($463$ pb$^{-1}$)}\\
\hline
\multicolumn{5}{c}{$M_{12}>$$100$~GeV}\\
\hline
Selection & Data & SM & Pair Production (GRAPE) & NC DIS + Compton \\
\hline
\multicolumn{5}{c}{All data ($463$ pb$^{-1}$)}\\
\hline                                        
$ee$  & $3$ & $1.34 \pm 0.20$ & $0.83 \pm 0.11$  & $0.51 \pm 0.13$ \\ 
$\mu\mu$  & $1$  & $0.17 \pm 0.07$ & $0.17 \pm 0.07$ &  $< 0.005$  \\ 
$e\mu$  & $1$   & $0.59 \pm 0.06$ & $0.59 \pm 0.06$ & $< 0.005$  \\ 
$eee$   & $3$ & $0.66 \pm 0.09$ & $0.66 \pm 0.09$ &  $< 0.005$    \\    
$e\mu\mu$   & $2$ & $0.16 \pm 0.05$ & $0.16 \pm 0.05$ &  $< 0.005$    \\
\hline
\multicolumn{5}{c}{$e^+p$ collisions ($285$ pb$^{-1}$)}\\
\hline                                        
$ee$  & $3$ & $0.76 \pm 0.11$ & $0.49 \pm 0.07$  & $0.27 \pm 0.07$ \\ 
$\mu\mu$  & $1$  & $0.10 \pm 0.04$ & $0.10 \pm 0.04$ &  $< 0.005$  \\ 
$e\mu$  & $1$   & $0.35 \pm 0.04$ & $0.35 \pm 0.04$ & $< 0.005$  \\ 
$eee$   & $3$ & $0.39 \pm 0.05$ & $0.39 \pm 0.05$ &  $< 0.005$    \\    
$e\mu\mu$ & $2$ & $0.09 \pm 0.03$ & $0.09 \pm 0.03$ &  $< 0.005$    \\
\hline                                        
\multicolumn{5}{c}{$e^-p$ collisions ($178$ pb$^{-1}$)}\\
\hline                                        
$ee$  & $0$ & $0.58 \pm 0.09$ & $0.34 \pm 0.04$ & $0.24 \pm 0.07$ \\ 
$\mu\mu$  & $0$  & $0.07 \pm 0.03$ & $0.07 \pm 0.03$ &  $< 0.005$  \\ 
$e\mu$  & $0$   & $0.24 \pm 0.03$ & $0.24 \pm 0.03$  & $< 0.005$  \\ 
$eee$   & $0$ & $0.27 \pm 0.04$ & $0.27 \pm 0.04$ &  $< 0.005$    \\    
$e\mu\mu$  & $0$ & $0.07 \pm 0.03$ & $0.07 \pm 0.03$ &  $< 0.005$    \\
\hline                                        
%
\end{tabular}
\end{center}
\caption{Observed and predicted multi-lepton event yields for masses $M_{12} > 100$~GeV for the different event classes in all analysed samples. 
  The errors on the predictions include model uncertainties and experimental systematic errors added in quadrature. The limits on the background estimations correspond to the selection of no event in the simulated sample and are quoted at $95$\% confidence level.}
\label{tab:mlepyieldsM100}
\end{table}

\begin{table}[]
\begin{center}
\begin{tabular}{ c c c c c }
\multicolumn{5}{c}{Multi-Leptons at HERA ($463$ pb$^{-1}$)}\\
\hline
\multicolumn{5}{c}{$\sum P_T>$$100$ GeV}\\
\hline
Data sample & Data & SM & Pair Production (GRAPE) & NC DIS + Compton \\
\hline                                        
e$^{+}$p ($285$ pb$^{-1}$)  & $5$ & $0.96 \pm 0.12$ & $0.78 \pm 0.09$  & $0.18 \pm 0.05$ \\ 
e$^{-}$p ($178$ pb$^{-1}$)  & $0$ & $0.64 \pm 0.09$ & $0.51 \pm 0.07$  & $0.13 \pm 0.04$ \\ 
All      ($463$ pb$^{-1}$)  & $5$ & $1.60 \pm 0.20$ & $1.29 \pm 0.15$  & $0.31 \pm 0.09$ \\ 
\hline                                        
\end{tabular}
\end{center}
\caption{Observed and predicted multi-lepton event yields for $\sum P_T >$ $100$~GeV. Di-lepton and tri-lepton events are combined.
  The errors on the predictions include model uncertainties and experimental systematic errors added in quadrature.}
\label{tab:mlepyieldsEt100}
\end{table}

%
\begin{center}
 \renewcommand{\arraystretch}{1.15} 
 \begin{table}[]
\begin{center}
\begin{tabular}{ l l l l l l l l l l l l  }
\multicolumn{12}{c}{Multi-Leptons at HERA ($463$~pb$^{-1}$)}\\
\hline
\multicolumn{1}{c}{Variable}  &  \multicolumn{3}{c}{Measured} & \multicolumn{3}{c}{Measured}        &    \multicolumn{3}{c}{Measured}  & \multicolumn{2}{c}{Pair Production} \\
\multicolumn{1}{c}{ range}    &   \multicolumn{3}{c}{  ($e^+e^-$) }  &   \multicolumn{3}{c}{ ($\mu^+\mu^-$)}  &    \multicolumn{3}{c}{(average)}  &  \multicolumn{2}{c}{ (GRAPE)}     \\
\multicolumn{1}{c}{\mbox{[GeV]}} & \multicolumn{3}{c}{ [fb/GeV] }  &  \multicolumn{3}{c}{[fb/GeV] }  &   \multicolumn{3}{c}{[fb/GeV]}    &   \multicolumn{2}{c}{[fb/GeV]}   \\
\hline
\multicolumn{1}{c}{$P_T^{\ell_1}$} & \multicolumn{10}{c}{ $d\sigma/dP_T^{\ell_1}$}\\
\hline
$[10,15]$ & ~~~~$108$&$\!\!\!\!\!\pm \, 10$&$\!\!\!\!\!\pm \, 8$&~~~~$95$&$\!\!\!\!\!\pm \, 9$&$\!\!\!\!\!\pm \, 14$&~~~~$101$&$\!\!\!\!\!\pm \, 7$&$\!\!\!\!\!\pm \, 9$&~~~~$92$&$\!\!\!\!\!\pm \, 3$ \\
$[15,20]$ & ~~~~$16$&$\!\!\!\!\!\pm \, 4$&$\!\!\!\!\!\pm \, 2$&~~~~$23$&$\!\!\!\!\!\pm \, 4$&$\!\!\!\!\!\pm \, 4$&~~~~$20$&$\!\!\!\!\!\pm \, 3$&$\!\!\!\!\!\pm \, 2$&~~~~$22$&$\!\!\!\!\!\pm \, 1$ \\
$[20,25]$ & ~~~~$7.4$&$\!\!\!\!\!\pm \, 2.5$&$\!\!\!\!\!\pm \, 0.6$&~~~~$3.7$&$\!\!\!\!\!\pm \, 1.7$&$\!\!\!\!\!\pm \, 0.5$&~~~~$5.5$&$\!\!\!\!\!\pm \, 1.5$&$\!\!\!\!\!\pm \, 0.5$&~~~~$6.6$&$\!\!\!\!\!\pm \, 0.2$ \\ 
$[25,50]$ & ~~~~$0.60$&$\!\!\!\!\!\pm \, 0.30$&$\!\!\!\!\!\pm \, 0.06$&~~~~$0.70$&$\!\!\!\!\!\pm \, 0.30$&$\!\!\!\!\!\pm \, 0.10$&~~~~$0.70$&$\!\!\!\!\!\pm \, 0.20$&$\!\!\!\!\!\pm \, 0.07$&~~~~$0.86$&$\!\!\!\!\!\pm \, 0.03$ \\ %
\hline
\multicolumn{1}{c}{$M_{\ell\ell}$}  & \multicolumn{10}{c}{ $d\sigma/dM_{\ell\ell}$}\\
\hline
$[15,25]$   & ~~~~$34$&$\!\!\!\!\!\pm \, 4$&$\!\!\!\!\!\pm \, 3$&~~~~$29$&$\!\!\!\!\!\pm \, 3$&$\!\!\!\!\!\pm \, 4$&~~~~$32$&$\!\!\!\!\!\pm \, 3$&$\!\!\!\!\!\pm  \,3$&~~~~$28$&$\!\!\!\!\!\pm  \,1$ \\
$[25,40]$   & ~~~~$16$&$\!\!\!\!\!\pm  \,2$&$\!\!\!\!\!\pm \, 2$&~~~~$17$&$\!\!\!\!\!\pm \, 2$&$\!\!\!\!\!\pm  \,3$&~~~~$17$&$\!\!\!\!\!\pm \, 2$&$\!\!\!\!\!\pm  \,2$&~~~~$18.0$&$\!\!\!\!\!\pm \, 0.5$ \\
$[40,60]$   & ~~~~$2.7$&$\!\!\!\!\!\pm \, 0.8$&$\!\!\!\!\!\pm \, 0.2$&~~~~$1.7$&$\!\!\!\!\!\pm \, 0.6$&$\!\!\!\!\!\pm \, 0.3$&~~~~$2.2$&$\!\!\!\!\!\pm \, 0.5$&$\!\!\!\!\!\pm \, 0.2$&~~~~$2.9$&$\!\!\!\!\!\pm \, 0.1$ \\ 
$[60,100]$  & ~~~~$0.20$&$\!\!\!\!\!\pm \, 0.15$&$\!\!\!\!\!\pm \, 0.02$&~~~~$0.40$&$\!\!\!\!\!\pm \, 0.20$&$\!\!\!\!\!\pm \, 0.07$&~~~~$0.30$&$\!\!\!\!\!\pm \, 0.13$&$\!\!\!\!\!\pm \, 0.04$&~~~~$0.24$&$\!\!\!\!\!\pm  \,0.01$ \\ 
\hline
\multicolumn{1}{c}{$P_T^X$}  & \multicolumn{10}{c}{ $d\sigma/dP_T^X$}\\
\hline
$[0,5]$   & ~~~~$95$&$\!\!\!\!\!\pm \, 9$&$\!\!\!\!\!\pm \, 8$&~~~~$94$&$\!\!\!\!\!\pm \, 8$&$\!\!\!\!\!\pm \, 14$&~~~~$94$&$\!\!\!\!\!\pm \, 6$&$\!\!\!\!\!\pm \, 9$&~~~~$93$&$\!\!\!\!\!\pm \, 3$ \\
$[5,12]$  & ~~~~$25$&$\!\!\!\!\!\pm \, 6$&$\!\!\!\!\!\pm \, 2$&~~~~$13$&$\!\!\!\!\!\pm \, 4$&$\!\!\!\!\!\pm \, 2$&~~~~$18$&$\!\!\!\!\!\pm \, 3$&$\!\!\!\!\!\pm \, 2$&~~~~$15.4$&$\!\!\!\!\!\pm \, 0.5$ \\
$[12,25]$ & ~~~~$2.8$&$\!\!\!\!\!\pm \, 1.3$&$\!\!\!\!\!\pm \, 0.3$&~~~~$4.2$&$\!\!\!\!\!\pm \, 1.2$&$\!\!\!\!\!\pm \, 0.6$&~~~~$3.8$&$\!\!\!\!\!\pm \, 0.9$&$\!\!\!\!\!\pm \, 0.4$&~~~~$3.8$&$\!\!\!\!\!\pm \, 0.1$ \\ 
$[25,50]$ & ~~~~$0.20$&$\!\!\!\!\!\pm  \,0.30$&$\!\!\!\!\!\pm \, 0.03$&~~~~$0.50$&$\!\!\!\!\!\pm \, 0.30$&$\!\!\!\!\!\pm \, 0.09$&~~~~$0.40$&$\!\!\!\!\!\pm \, 0.20$&$\!\!\!\!\!\pm \, 0.06$&~~~~$0.20$&$\!\!\!\!\!\pm \, 0.01$ \\ 
\hline                                        
\end{tabular}
\end{center}
\caption{Differential cross sections $d\sigma/dP_T^{\ell_1}$, $d\sigma/dM_{\ell\ell}$ and $d\sigma/dP_T^X$ averaged for each quoted interval for the process $ep \rightarrow e \ell^+ \ell^- X$ in a restricted phase space dominated by the photon-photon process (see text for details). 
Cross sections are measured for $e^+e^-$ or $\mu^+\mu^-$ pairs.
The cross section obtained from the combination of $e^+e^-$ and $\mu^+\mu^-$ channels is also presented.
The first error is statistical and the second is systematic.
Theoretical predictions, calculated with GRAPE, are shown in the last column.}
\label{tab:xsection}
\end{table}
\end{center}

\vfill
\newpage

\vfill
\newpage

\end{document}